\documentclass[cits]{PoS}
\usepackage{amsmath,amssymb}
\usepackage{graphicx}
\newcommand{\ssst}{\scriptscriptstyle}

\newcommand{\UC}{\mathbf{1}_c}
\newcommand{\Rx}{R_{\mathbf{x}}}
\title{
Meson mass spectrum
at finite temperature and density
in the strong coupling limit of lattice QCD
for color SU(3)
}

\ShortTitle{Meson mass in the SC-LQCD at finite $T$ and $\mu$}
\author{Noboru Kawamoto, \speaker{Kohtaroh Miura} and Akira Ohnishi\\
Hokkaido University\\
E-mail: \email{miura@particle.sci.hokudai.ac.jp}}
\abstract{
We investigate the meson mass spectrum
in the strong coupling limit of lattice QCD
with one species of staggered fermion
for the SU($N_c$) color gauge group,
including $N_c=3$.
We analytically derive meson masses
as functions of temperature and chemical potential via chiral condensates. 
We show that meson masses quickly decrease to zero when the chemical 
potential or the temperature approaches to the critical value. 
}
\FullConference{The XXV International Symposium on Lattice Field Theory\\
		 July 30-4 August 2007\\
		 Regensburg, Germany}
\begin{document}
\section{Introduction}
It is a common folklore that the chiral phase transition takes place
at high temperature and/or density
in Quantum Chromodynamics (QCD).
Meson masses are crucially influenced by the chiral condensate,
and thus are
very interesting observables 
in the chiral phase transition region.

It is possible to extract hadron masses quantitatively
in the lattice Monte-Carlo (MC) simulations at zero density.
However, MC does not work well 
at high densities 
because of the notorious negative sign problem of the Dirac determinant.
Therefore it is instructive to
investigate high density matter
in the strong coupling lattice QCD (SC-LQCD).
In SC-LQCD analyses with a mean field approximation (MFA), 
we can derive analytical expressions of the effective potential
as a function of temperature ($T$) and chemical potential ($\mu$),
and hence we can avoid the sign problem.
The numerical values of physical observables derived analytically 
in SC-LQCD should be reproduced
in lattice MC simulations at least in the strong coupling region. 
From this point of view,
Refs.~\cite{MC_SCL} provides an interesting comparison of
strong coupling results in MC and analytic studies.

Chiral phase transitions are governed by the effective potential,
which also plays an essential role to obtain the hadron mass in SC-LQCD.
The effective potential at finite temperature and density
has been extensively studied
under various 
conditions~\cite{DKS,Faldt1986,Nishida:2003fb,SCL_Feff,Damgaard:1985bn,OurPaper}.
For example, we have recently derived
analytical expressions of the effective potential
as a function of $T$ and $\mu$ in SC-LQCD,
including the baryon propagating effects 
or $1/g^2$ corrections~\cite{OurPaper}.
On the other hand, while hadron masses have been studied 
at zero temperature~\cite{Damgaard:1985bn,SCL_Hm,Kluberg-Stern:1982bs},
the expression of meson masses as a function of $T$ and $\mu$
has never been derived before.

In this proceedings, 
we derive an analytical expression of meson masses
as functions of the chiral condensate,
which is a function of $T$ and $\mu$ in SC-LQCD for color SU($N_c$).

\section{A brief summary of meson mass derivation}
We start from the action and partition function of
lattice QCD with one species of staggered fermion($\chi$)
in the strong coupling limit,
where we omit the pure gauge plaquette terms ($\propto 1/g^2$),  
\begin{eqnarray}
Z&=&
\int_{\chi,\bar{\chi},U_0,U_j}\exp
\Bigl[
-\frac{1}{2}\sum_{x}\sum_{j=1}^d
(-1)^{x_0+\cdots+x_j}\bigl[
\bar{\chi}_{x}U_j(x)\chi_{x+\hat{j}}-(h.c)
\bigr]
-S_{\ssst F}^{~t}-m_0\sum_xM_x)
\Bigr]\\
S_{\ssst F}^{~t}
&=&
\frac{1}{2}\sum_{x}
\bigl[
e^{\mu}\bar{\chi}_xU_0(x)\chi_{x+\hat{0}}-e^{-\mu}(h.c)
\bigr]
\ ,\quad
M_x=\sum_{a=1}^{N_c}(\bar{\chi}^a\chi_a)_x
\ ,
\end{eqnarray}
where $m_0,~U_0,~U_j$ represent
the current quark mass, the temporal and spatial link variables,
respectively.
We introduce the lattice chemical potential $\mu$ 
following the procedure in Ref.~\cite{Hasenfratz:1983ba}.

First we perform the path integral over spatial link variables $U_j$,
and keep only the leading order terms in the $1/d$ expansion,
\begin{eqnarray}
Z=\int_{\chi,\bar{\chi},U_0}\exp
\Bigl[
-S_{\ssst F}^{~t}
+\frac{\sum_{x,j}M_xM_{x+\hat{j}}}{4N_c}
-m_0\sum_xM_x
+\mathcal{O}(1/\sqrt{d})
\Bigr]
\ .
\end{eqnarray}
Next we introduce 
the chiral condensate $\sigma \propto \langle\chi\bar{\chi}\rangle$
through the bosonization procedure,
\begin{eqnarray}
Z&\simeq&
\int_{\chi,\bar{\chi},U_0,\sigma}\exp
\Bigl[
-\frac{1}{2}\sum_{mn,\mathbf{xy}}
\sigma_{m\mathbf{x}}
V_{\ssst M}^{-1}(\mathbf{xy})\delta_{mn}
\sigma_{n\mathbf{y}}
-\sum_{n\mathbf{x}}m^{(q)}_{n\mathbf{x}}M_{n\mathbf{x}}
-S_{\ssst F}^{~t}
\Bigr]
\ ,\label{eq:ZHS}\\
V_{\ssst M}(\mathbf{xy})
&=&
\frac{1}{4N_c}
\sum_{j=1}^{d}
(\delta_{\mathbf{x}+\hat{j},\mathbf{y}}+\delta_{\mathbf{x}-\hat{j},\mathbf{y}})
\ ,\quad
m^{(q)}_{n\mathbf{x}}=m_0+\sigma_{n\mathbf{x}}
\ ,
\end{eqnarray}
where ``$m$'' or ``$n$'' denotes the temporal lattice site,
which takes an integer value in $[1,N]$.
Now we can perform 
the path integral over staggered quarks,
and obtain the following effective action of $\sigma$ up to a constant,
\begin{equation}
Z\simeq
\int\mathcal{D}{\sigma}~e^{-S_{\text{eff}}[\sigma]}
\ ,\quad
S_{\text{eff}}[\sigma]
=\frac{1}{2}\sum_{mn,\mathbf{xy}}
\sigma_{m\mathbf{x}}
V_{\ssst M}^{-1}(\mathbf{xy})\delta_{mn}
\sigma_{n\mathbf{y}}
-\sum_{\mathbf{x}}\log \Rx[\sigma]
\ .\label{eq:Seff}
\end{equation}
Integral over $\chi$ and $U_0$ generates
the interaction term\ $\log\,\Rx[\sigma]$,
where $\Rx[\sigma]$ is given as,
\begin{equation}
\Rx[\sigma]
=
\int dU_0(\mathbf{x})
\begin{vmatrix}
2m^{(q)}_{1,\mathbf{x}}\UC&e^{\mu}\UC&&&e^{-\mu}U_0^{\dagger}(\mathbf{x})\\
-e^{-\mu}\UC&2m^{(q)}_{2,\mathbf{x}}\UC&e^{\mu}\UC&\textbf{\Large{0}}&\\
&\ddots&\ddots&\ddots&\\
&\textbf{\Large{0}}&-e^{-\mu}\UC&2m^{(q)}_{N-1,\mathbf{x}}\UC&e^{\mu}\UC\\
-e^{\mu}U_0(\mathbf{x})&&&-e^{-\mu}\UC&2m^{(q)}_{N,\mathbf{x}}\UC
\end{vmatrix}
\ ,\label{eq:R}
\end{equation}
where $\UC$ represents an $N_c\times N_c$ unit matrix,
and $U_0$
is given in the temporal gauge as,
\begin{equation}
U_0(\mathbf{x})=
\mathrm{diag}
\{e^{i\theta^1(\mathbf{x})},\cdots,e^{i\theta^{N_c}(\mathbf{x})}\}
\ .
\end{equation}
%
Periodic and anti-periodic boundary conditions for gluons and fermions
are respected in the present finite $T$ treatment
as found in the form of Eq.~(\ref{eq:R}),
where we find different signs
in the upper-right and the lower-left components.

We decompose the chiral condensates $\sigma_n(\mathbf{x})$
to the equilibrium value $\bar{\sigma}$ 
and fluctuations $\delta\sigma_n(\mathbf{x})$,
$\sigma_n(\mathbf{x})=\bar{\sigma}+\delta\sigma_n(\mathbf{x})$.
The equilibrium value $\bar{\sigma}$ is determined
from the stationary condition of the effective action, 
\begin{equation}
\frac{\delta S_{\text{eff}}[\sigma]}
{\delta \sigma_{n\mathbf{x}}}
\Big|_{\sigma\to\bar{\sigma}}
=
\bar{\sigma}\frac{2N_c}{d}
-\frac{1}{R[\bar{\sigma}]}
\frac{\delta R_{\mathbf{x}}[\sigma]}
{\delta \sigma_{n\mathbf{x}}}
\Big|_{\sigma\to\bar{\sigma}}
=0
\ .
\label{eq:delR}
\end{equation}
The meson mass is defined as the pole of 
the propagator for $\delta\sigma$,
which is obtained from 
the second order variation of the effective action
($\delta^2_{\sigma}S_{\text{eff}}$).
The stationary condition Eq.~(\ref{eq:delR}) ensures that
the first derivative of $R_{\mathbf{x}}[\sigma]$
is independent from the space-time point.
Thus when we require the null average fluctuation condition
$\sum_n\delta\sigma_n=0$,
the first derivative of $R_{\mathbf{x}}[\sigma]$ 
in $\delta^2_{\sigma}S_{\text{eff}}$ disappears.
The second order variation $\delta^2_{\sigma}S_{\text{eff}}$
is found to be,
\begin{equation}
\delta_{\sigma}^2 S_{\text{eff}}[\sigma]\Big|_{\sigma\to\bar{\sigma}}
=\sum_{mn,\mathbf{xy}}
\delta\sigma_{m\mathbf{x}}
\Biggl[
V_{\ssst M}^{-1}(\mathbf{xy})\delta_{mn}
-\delta_{\mathbf{xy}}
\frac{R^{(2)}_{mn}[\sigma]}
{R[\sigma]}
\Bigg|_{\sigma\to\bar{\sigma}}
\Biggr]
\delta\sigma_{n\mathbf{y}}
\label{eq:del2S}\ ,\quad
R^{(2)}_{mn}
\equiv\frac{\delta^2 R_{\mathbf{x}}}{\delta\sigma_m\delta\sigma_n}
\ .
\end{equation}
It is necessary to evaluate $R^{(2)}_{mn}$
to obtain meson masses.
We derive $R^{(2)}_{mn}$ at finite $T$ and $\mu$
in the following section, which has not been derived previously.

\section{Evaluation of quark hopping in the temporal direction}
We utilize the formulation developed
by Damgaard, Kawamoto, Shigemoto~\cite{DKS},
Faldt and Petersson~\cite{Faldt1986},
and Nishida~\cite{Nishida:2003fb}.
First we reduce the
$(N_c\times N)\times(N_c\times N)$
determinant in Eq.~(\ref{eq:R})
to 
$N_c\times N_c$
determinant in the form~\cite{Faldt1986},
\begin{equation}
R_{\mathbf{x}}
=\int dU_0
\det_c\Bigl[
X_N\otimes\UC
+(e^{\mu/T}U_0+e^{-\mu/T}U_0^{\dagger})
\Bigr]
=R_{\mathbf{x}}(X_N[\sigma],\mu)
\label{eq:decompR}
\ ,
\end{equation}
where the quark hopping kernel $X_N$ is a functional of $\sigma$.
We assume that the number of temporal lattice sites $N$
is even and its inverse $N^{-1}$ is identified as the temperature $T$.
%
Since $\Rx$ is a function of $X_N$,
it is enough to evaluate the $U_0$ integral in equilibrium,
and this has been done by utilizing the Vandermonde determinant technique
combined with the recursion formula~\cite{DKS,Nishida:2003fb}.
It is also possible to perform the $U_0$ integral explicitly
by using the one link integral technique~\cite{Faldt1986}.
%
The obtained $\Rx$ reads,
\begin{equation}
R_{\mathbf{x}}(X_N[\sigma]=Y+Y^{-1},\mu)
=\frac{Y^{N_c+1}-Y^{-(N_c+1)}}
{Y-Y^{-1}}
+2\mathrm{cosh}\frac{N_c\mu}{T}
\ .\label{eq:RY}
\end{equation}
In this proceedings, 
the baryonic effects are considered for just a temporal direction,
which is reflected on the second term in Eq~(\ref{eq:RY}).

Next it is necessary to evaluate 
the quark hopping kernel $X_N$, which is found to be~\cite{Faldt1986},
\begin{equation}
X_N=B_{1,\cdots,N}+B_{2,\cdots,N-1}\ ,
\quad
B_{1,\cdots,N}
=
\begin{vmatrix}
2m^{(q)}_{1,\mathbf{x}}&e^{\mu}&&&\\
-e^{-\mu}&2m^{(q)}_{2,\mathbf{x}}&e^{\mu}&&\textbf{\Large{0}}\\
&\ddots&\ddots&\ddots&\\
\textbf{\Large{0}}&&-e^{-\mu}&2m^{(q)}_{N-1,\mathbf{x}}&e^{\mu}\\
&&&-e^{-\mu}&2m^{(q)}_{N,\mathbf{x}}
\end{vmatrix}
\label{eq:B}
\ .
\end{equation}
In equilibrium 
($\delta\sigma_n(\mathbf{x})\to 0,~\sigma_n(\mathbf{x})\to\bar{\sigma}$),
recursion relation 
$B_n = 2m^{(q)}B_{n-1}+B_{n-2}$ 
leads to the equilibrium value of $Y$ and $B$ as,
\begin{eqnarray}
\bar{Y}&=&Y[\sigma\to\bar{\sigma}]=e^{E/T}\ ,\label{eq:barY}\\
\bar{B}_n\equiv B_{k,\cdots,k+n}(\sigma\to\bar{\sigma})
&=&
\begin{cases}
\mathrm{cosh}
\bigl[
(n+1)E
\bigr]
/\mathrm{cosh} E\quad
(n=\text{even})\\
\mathrm{sinh}
\bigl[
(n+1)E
\bigr]
/\mathrm{cosh} E\quad
(n=\text{odd})
\end{cases}
\label{eq:barB}
\ ,
\end{eqnarray}
where $E=\mathrm{sinh}^{-1}(m_0+\bar{\sigma})$ denotes
the one dimensional quark energy.
By substituting $\bar{Y}$ in $\Rx$, 
we get the effective potential from the effective action (Eq.~(\ref{eq:Seff})),
\begin{equation}
\mathcal{F}_{\text{eff}}(\bar{\sigma})
=S_{\text{eff}}(\bar{\sigma})/\sum_x
=\frac{1}{2}
\frac{2N_c}{d}\bar{\sigma}^2
-T\log\Bigl[
\frac{\mathrm{sinh}[(N_c+1)E(\bar{\sigma})/T]}
{\mathrm{sinh}[E(\bar{\sigma})/T]}+2\mathrm{cosh}\frac{N_c\mu}{T}
\Bigr]
\ .\label{eq:Feff}
\end{equation}
This has been obtained also in Ref.~\cite{Nishida:2003fb}
and included in the effective potential
in Ref.~\cite{OurPaper}.
The critical values $T_c(\mu=0),~\mu_c(T=0)$ and
the tri-critical point $T_{\text{tcp}}$ can be extracted 
from the effective potential~Eq.~(\ref{eq:Feff})
and have been studied
in Ref.~\cite{Nishida:2003fb} in the chiral limit,
\begin{eqnarray}
T_c(\mu=0)&=&d(N_c+1)(N_c+2)/\bigl[6(N_c+3)\bigr]\ ,\quad
\mu_c(T=0)\simeq 0.55
\label{eq:cri}\\
T_{\text{tcp}}
&=&
\bigl[
\sqrt{225N_c^2+20d^2(3N_c^2+6N_c-4)}-15N_c
\bigr]/(20d)
\ .\label{eq:Ttcp}
\end{eqnarray}

Finally, it is necessary to evaluate $R^{(2)}_{mn}$
to obtain the meson mass spectrum(See Eq.~(\ref{eq:del2S})).
Here, the translational invariance simplifies the calculation.
Since the system has translational invariance, 
$\Rx$ and $X_N$ are cyclic invariant for the temporal indices.
By using this translational invariance we obtain,
\begin{eqnarray}
\left.\frac{\partial X_N}{\partial\sigma_n}
\right|_{\bar{\sigma}}
&=&
\left.\frac{\partial X_N}{\partial\sigma_{\ssst N}}
\right|_{\bar{\sigma}}
=2\bar{B}_{N-1}
\label{eq:Xn}\\
\left.\frac{\partial^2 X_N}{\partial\sigma_m\partial\sigma_n}\right|_{\bar{\sigma}}
&=&
\left.
\frac{\partial^2 X_N}
{\partial\sigma_{n^{\prime}}\partial\sigma_{\ssst N}}\right|_{\bar{\sigma}}
=
4\bar{B}_{n^{\prime}-1}\bar{B}_{N-n^{\prime}-1}
\quad (n^{\prime}=N-|n-m|)
\ .\label{eq:Xmn}
\end{eqnarray}
It is found that
Eqs.~(\ref{eq:RY}), (\ref{eq:Xn}) and (\ref{eq:Xmn}) lead to,
\begin{equation}
R^{(2)}_{mn}
\Big|_{\bar{\sigma}}
=
4(\bar{B}_{N-1})^2
{\partial^2R\over\partial X_N^2}
\Big|_{\bar{\sigma}}
+4\bar{B}_{n'-1}\bar{B}_{N-n'-1}
{\partial R\over\partial X_N}
\Big|_{\bar{\sigma}}
\ ,\label{eq:Rmn}
\end{equation}
in a straightforward calculation.
In equilibrium,
we can evaluate $(dR/dX_N)|_{\bar{\sigma}}$ in the following way,
\begin{equation}
\frac{1}{\bar{R}}\frac{dR}{dX_N}\Big|_{\bar{\sigma}}
=\frac{1}{\bar{R}}
\frac{\partial R}{\partial\sigma_n}
\Bigl[\frac{\partial X_N}{\partial\sigma_n}\Bigr]^{-1}\Big|_{\bar{\sigma}}
=\bar{\sigma}\frac{2N_c}{d}\frac{1}
{2\bar{B}_{N-1}}
\ ,\label{eq:dR/barR}
\end{equation}
where we used Eq.~(\ref{eq:delR}) and (\ref{eq:barY})
in the final
equality.
Taking into account the even number of temporal lattice cite ($N=$even),
and substituting Eq.~(\ref{eq:barB}) and (\ref{eq:dR/barR})
into (\ref{eq:Rmn}),
we obtain,
\begin{equation}
\frac{\bar{R}_{mn}^{(2)}}{\bar{R}}
=-\bar{\sigma}\frac{2N_c}{d}
\frac{e^{i\pi m^{\prime}}
\mathrm{cosh}\bigl[(N-2m^{\prime})E\bigr]}
{\mathrm{cosh} E~\mathrm{sinh}[E/T]}
+\frac{\Delta\mathcal{R}}{\bar{R}}
\quad (m^{\prime}=|n-m|)\label{eq:Rm/Rbar}\ ,
\end{equation}
where $\Delta\mathcal{R}$ does not include temporal indices,
and here we do not bother to write its explicit form.
\section{Meson mass}
Now we can explicitly evaluate t
he second variation of the effective action~(Eq.~(\ref{eq:del2S})).
Substituting Eq.~(\ref{eq:Rm/Rbar}) into (\ref{eq:del2S}),
we obtain,
\begin{equation}
\delta^2_{\sigma} 
S_{\text{eff}}
\Big|_{\bar{\sigma}}
=
\sum_{mn,\mathbf{xy}}
\delta\sigma_{m\mathbf{x}}
\Biggl[
V_{\ssst M}^{-1}(\mathbf{xy})\delta_{mn}
+\delta_{\mathbf{xy}}
\bar{\sigma}\frac{2N_c}{d}
\frac{e^{i\pi m^{\prime}}\mathrm{cosh}\bigl[(N-2m^{\prime})E\bigr]}
{\mathrm{cosh} E~\mathrm{sinh}[E/T]}
\Biggr]
\delta\sigma_{n\mathbf{y}}
\ ,\label{eq:del2S_2}
\end{equation}
where we ignore 
the effect of $\Delta\mathcal{R}/\bar{R}$
by requiring the null average condition: 
$\sum_n\delta\sigma_n=0$. 
We perform the Fourier transformation of Eq.~(\ref{eq:del2S_2})
by using the translational invariance,
$\sum_{n=1}^{N}
\equiv\sum_{m^{\prime}=1-m}^{N-m}
\to \sum_{m^{\prime}=0}^{N-1}$
for the second term in Eq.~(\ref{eq:del2S_2}), 
and we obtain,
\begin{equation}
\delta^2_{\sigma} 
S_{\text{eff}}
\Big|_{\bar{\sigma}}
=
\sum_{\omega,\mathbf{k}}
\delta\sigma_{\omega}(\mathbf{k})
\Biggl[
\frac{2N_c/d}{\sum_j\cos{k}_j}
+\frac{(2N_c/d)\bar{\sigma}(\bar{\sigma}+m_0)}
{\cos\omega+2(\bar{\sigma}+m_0)^2+1}
\Biggr]
\delta\sigma_{\omega}(\mathbf{k})
\ .\label{eq:s3final}
\end{equation}
Here we introduce the prescription 
proposed in Ref.~\cite{Kluberg-Stern:1982bs},
\begin{equation}
(\omega,\mathbf{k})
=
(iM,\mathbf{0})+(\delta_{\nu})\pi
\ ,\label{eq:decomp}
\end{equation}
where $(\delta_{\nu})$ is a ``$d+1$'' dimensional vector
which takes $0$ or $1$ and originates from the tastes degrees of freedom.
In this prescription, 
``$M$'' is regarded as a meson mass.
By putting the obtained inverse propagator for
$\delta\sigma_{\omega}(\mathbf{k})$ equal to zero,
the meson masses are found to be,
\begin{eqnarray}
\pm\mathrm{cosh} M_{\kappa}(\bar{\sigma};T,\mu)
&=&
2(\bar{\sigma}+m_0)
\Bigl(
\frac{d+\kappa}{d}\bar{\sigma}+m_0
\Bigr)+1\label{eq:Mm}\\
\kappa
&\equiv&
\sum_{j=1}^d \cos \delta_j
\in \{-d,-d+2,\cdots,d-2,d\}
\ .
\end{eqnarray}
Here, we tentatively take the plus sign
to obtain the real number meson mass,
and regard taste effects $\kappa$ as meson species
in the same way as in Ref.~\cite{Kluberg-Stern:1982bs}.
In this scheme with $d=3$, ``$\kappa =-3,~-1,~1,~3$''
corresponds to ``$\pi,~\rho,~a_1$'' 
and ``$\delta$'' mesons, respectively.

\begin{figure}
\begin{center}
\scalebox{1.1}[1.1]{\includegraphics{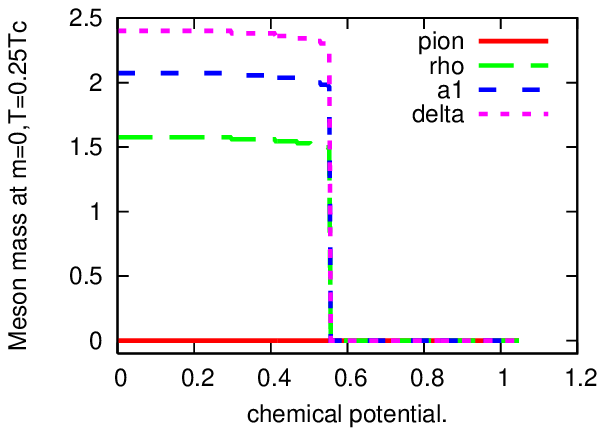}}
\scalebox{1.1}[1.1]{\includegraphics{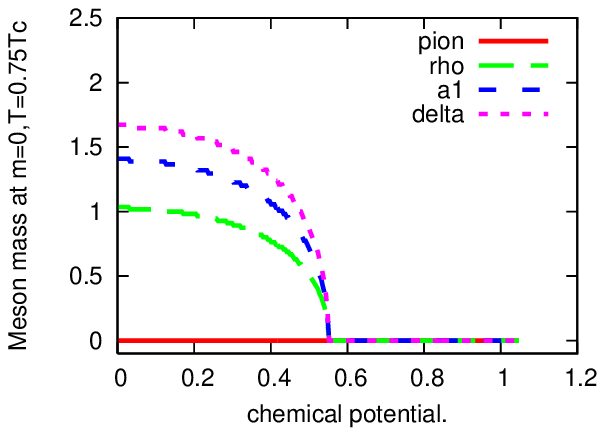}}
\scalebox{1.1}[1.1]{\includegraphics{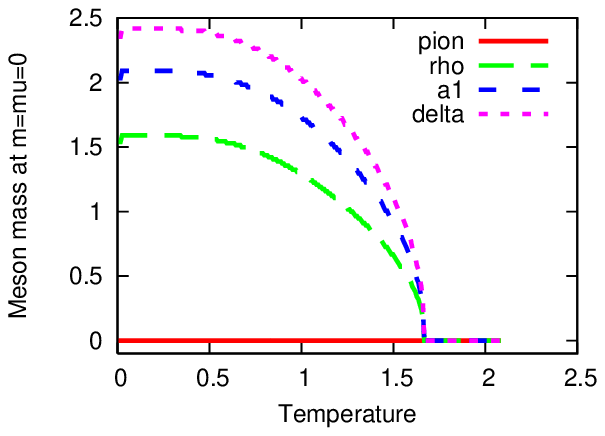}}
\scalebox{1.1}[1.1]{\includegraphics{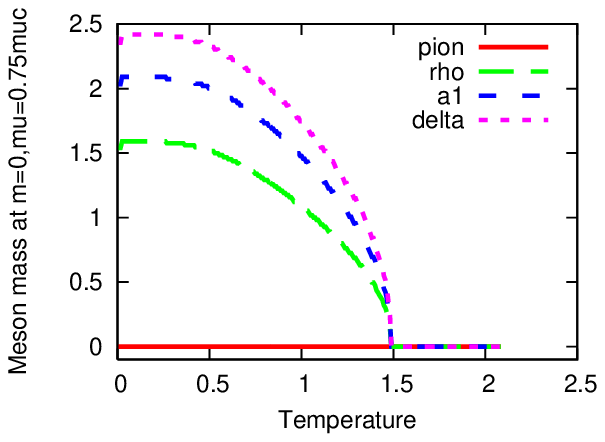}}
\caption{The $\mu$ dependence
of meson mass at $T=0.25T_c(\mu=0)$(:upper-left) and
 $T=0.75T_c(\mu=0)$(:upper-right). $T$ dependence of meson
 mass at $\mu=0$(:lower-left) and
 $\mu=0.75\mu_c(T=0)$(:lower-right).
 From \protect\ref{eq:cri}.
 We are considering the chiral limit
 case in the lattice unit.}\label{fig:Mm_0}
\end{center}
\end{figure}

For a small current quark mass $m_0\sim 0$,
we obtain, $M_{\pi}\simeq 2\sqrt{\bar{\sigma}m_0}$,
which may be regarded as the PCAC relation.
When the temperature $T$ and quark chemical potential $\mu$
are given, the chiral condensate $\bar{\sigma}(T,\mu)$
is obtained from the effective potential Eq.~(\ref{eq:Feff}), 
then the meson mass $M_{\kappa}(\bar{\sigma};T,\mu)$ 
is determined by Eq.~(\ref{eq:Mm}).
In Fig.~\ref{fig:Mm_0},
we show $T$ and $\mu$ dependence
of meson masses in the lattice unit.
With a given $T$
lower than the tri-critical point
$T_{\text{tcp}}$ ({\it c.f.} Eq.~(\ref{eq:Ttcp})),
the meson masses except for pion discontinuously
drop when $\mu$ passes through 
the critical value $\mu_c$,
where the phase transition is first order
(upper-left panel of Fig.~\ref{fig:Mm_0}).
At higher temperatures, $T_{\text{tcp}}\leq T<T_c$,
the phase transition is second order and thus 
the meson masses except for pion quickly but smoothly
decrease to zero 
when $\mu$ approaches to $\mu_c$ 
(upper-right panel of Fig.~\ref{fig:Mm_0}).
When $T$ approaches to its critical value $T_c$ 
with a given $\mu$ lower than $\mu_c$,
the meson masses quickly and smoothly decrease to zero,
where the phase transition is second order 
(lower panels of Fig.~\ref{fig:Mm_0}).
The $T$ dependence of meson masses
is slightly affected by a given $\mu$
in the chiral broken phase~(lower panels of Fig.~\ref{fig:Mm_0}).

\section{Summary}
We have investigated the meson mass spectrum
at finite temperature and density
by considering the leading order of the $1/d$ expansion
in the strong coupling limit of lattice QCD.
We have derived an analytical expression of meson masses as functions of
the chiral condensate which is a function
of temperature and chemical potential.
We have thus explicitly studied 
the temperature and chemical potential dependence 
of meson masses near the critical value. We have confirmed
that the pion mass satisfies the PCAC relation,
and that masses of other mesons decrease to zero very quickly 
when the chemical potential or temperature approaches to
the critical value.
\section*{Acknowledgements}
This work is supported in part by the Ministry of Education,
Science, Sports and Culture,
Grant-in-Aid for Scientific Research
under the grant numbers,
    13135201,		
    15540243,		
    1707005,		
and 19540252.		

\end{document}